\documentclass[a4paper, superscriptaddress, aps, reprint, prl, onecolumn]{revtex4}
\usepackage{amsmath}
\usepackage{graphicx}
\usepackage[labelfont=bf,justification=centerlast,labelsep=endash]{caption}

\makeatletter 

\renewcommand{\thetable}{\@arabic\c@table}

\makeatother 

\begin{document}
\title{Interaction imaging with amplitude-dependence force spectroscopy}
\author{Daniel Platz}
\email{platz@kth.se}
\author{Daniel Forchheimer}
\affiliation{Royal Institute of Technology (KTH), Section for Nanostructure Physics, Albanova University Center, SE-106 91 Stockholm, Sweden}
\author{Erik A. Thol\'{e}n}
\affiliation{Intermodulation Products AB, SE-169 58 Solna, Sweden}
\author{David B. Haviland}
\affiliation{Royal Institute of Technology (KTH), Section for Nanostructure Physics, Albanova University Center, SE-106 91 Stockholm, Sweden}

\begin{abstract}
Knowledge of surface forces is the key to understanding a large number of
processes in fields ranging from physics to material science and biology. The
most common method to study surfaces is dynamic atomic force microscopy (AFM).
Dynamic AFM has been enormously successful in imaging surface topography, even
to atomic resolution, but the force between the AFM tip and the surface remains
unknown during imaging. Here, we present a new approach that combines high
accuracy force measurements and high resolution scanning. The method, called
amplitude-dependence force spectroscopy (ADFS) is based on the
amplitude-dependence of the cantilever's response near resonance and allows for
separate determination of both conservative and dissipative tip-surface
interactions. We use ADFS to quantitatively study and map the nano-mechanical
interaction between the AFM tip and heterogeneous polymer surfaces. ADFS is
compatible with commercial atomic force microscopes and we anticipate its
wide-spread use in taking AFM toward quantitative microscopy.
\end{abstract}

\maketitle

\section{Introduction}
High-quality-factor resonators are increasingly used in a wide variety of
ultra-sensitive measurements of force\cite{Mamin2001}, mass\cite{Jensen2008} and
motion\cite{LaHaye2004}. High quality factor means that the oscillator can
respond with large changes in oscillation amplitude or phase, to very small
changes in a perturbing force. The dominant measurement paradigm for exploiting
this enhanced sensitivity is based on driving the resonator with one pure tone
near its resonance, while monitoring the response to the perturbation as a
change of amplitude or phase, or even a frequency shift when a feedback loop is
used to lock the phase. In the field of AFM this measurement paradigm has been
used to create images of surfaces to atomic resolution\cite{Giessibl1995} with
very little back-action on soft and delicate material surfaces. However, in
spite of its name, the atomic 'force microscope' does not actually measure the
force between the tip and the surface while imaging in these modalities. The
determination of tip-surface force requires that one monitors the change in
response as one slowly changes the height of the resonator above the
surface\cite{Durig2000,Sader2005,Holscher2006,Lee2006,Hu2008,Katan2009}. Such
measurements have provided impressive three dimensional force-volume
plots\cite{Holscher2002,Baykara2010} but they are fundamentally limited by
extremely long acquisition times which necessitate enormous effort to reduce
drift artifacts\cite{Albers2009}, making three dimensional force maps
impractical for most applications.

In this article we demonstrate a new approach to dynamic interaction measurement
based on actively modulating the amplitude of oscillation while measuring the
amplitude-dependence of the resonators response. A variation of amplitude which
is slow on the time scale of the oscillation period lends itself to an analysis
of the perturbing force in terms of the amplitude dependence of two signals: one
in-phase and one quadrature to the sinusoidal motion. The amplitude-dependence
of these two force signals allows us to determine the tip-surface interaction
directly as a function of the cantilever deflection \emph{at fixed probe
height}. We show how these two quadrature force signals provide the foundation
for model-free determination of the conservative force versus distance curve, as
well as an energy loss versus amplitude curve. These two force signals represent
essentially all the information that is possible to obtain about the nonlinear
perturbing force, from analysis of oscillator motion in the limited frequency
band near the high $Q$ resonance. After deriving the theory of ADFS, we test the
concept both with numerical simulation and experimental study of heterogeneous
polymer surfaces. The improved force resolution of ADFS allows us to critically
analyze a commonly accepted model of the tip-surface interaction and it enables
an alternative approach for characterizing the mechanical properties of polymer
surfaces.

\section{Results}

\subsection{Dynamics and force modeling}
In air and in vacuum the continuum dynamics of a cantilever beam are usually
reduced to a single mode harmonic oscillator described by a simple equation of
motion\cite{Melcher2007}
\begin{equation}
\ddot{z}+\frac{\omega_{0}}{Q}\dot{z}+\omega_{0}^{2}(z-h)=\frac{\omega_{0}^{2}}{k}F_{\mathrm{drive}}(t)+\frac{\omega_{0}^{2}}{k}F_{\mathrm{ts}}\left(z,\dot{z},\{z(t)\}\right)\label{eq:equation-of-motion}
\end{equation}
where $z$ is the position of the tip in the lab frame, h corresponds to the
static probe height above the surface and $\omega_{0}=\sqrt{k/m}$ is the
frequency of the first flexural cantilever resonance with quality factor $Q$,
effective spring constant $k$ and effective mass $m$. The two forces acting on
the oscillator are the tip-surface force $F_{\mathrm{ts}}$ and the external
sinusoidal drive force of strength $F_{\mathrm{drive}}$. The tip-surface force
$F_{\mathrm{ts}}$ is in general a complicated function depending on the
instantaneous tip position $z(t)$, velocity $\dot{z}(t)$ and, for hysteretic
forces, on the past tip trajectory $\{z(t)\}$. One typically decomposes the
tip-surface interactions into an effective conservative part which only depends
on the tip position $z$ and a more complicated effective dissipative
part\cite{Sader2005},
\begin{equation}
F_{\mathrm{ts}}(z,\dot{z},\{z\})=F_{\mathrm{c}}(z)+F_{\mathrm{nc}}\left(z,\dot{z},\lbrace z(t)\rbrace\right)\label{eq:force-decomposition}
\end{equation}
When driving with a sinusoidal signal on resonance, $\omega_{\mathrm{drive}}=\omega_{0}$,
the tip motion $z(t)$ is approximately sinusoidal\cite{Cleveland1998},
\begin{equation}
z(t)\approx A\cos(\omega_{0}t+\varphi)+h,\label{eq:tip-motion}
\end{equation}
with amplitude $A$, frequency $\omega_{0}$ and phase lag $\varphi$ with respect
to the drive force, even under the influence of the significant tip-surface
interactions. This sinusoidal tip motion is a result of the large quality factor
($Q\approx100$ in air, $Q\approx10000$ in vacuum) which guarantees that the
energy in the cantilever oscillation is greater than the tip-surface interaction
energy during one oscillation cycle.

\begin{figure}[ht]
\includegraphics{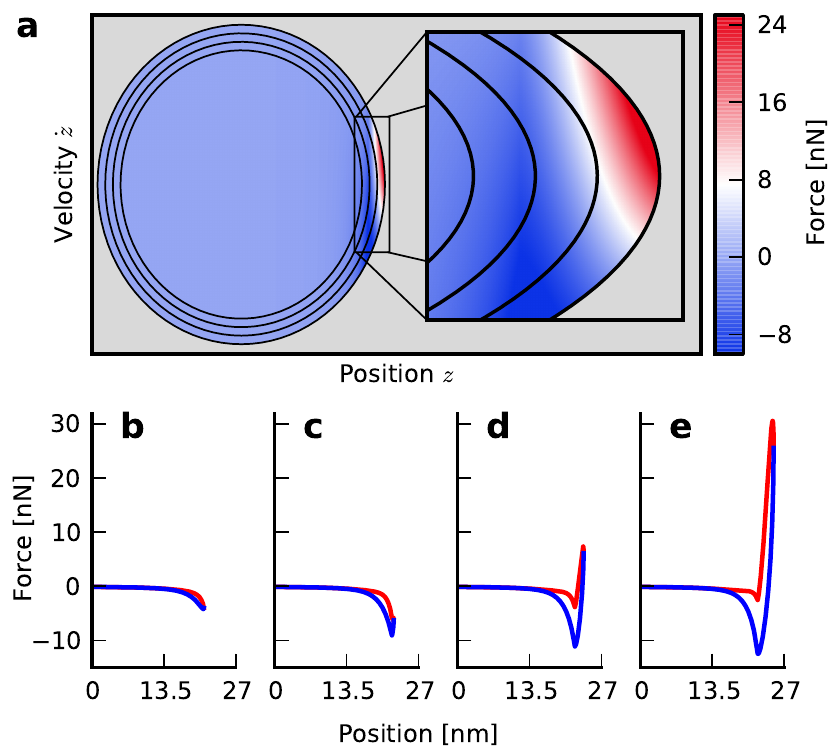}
\caption{\textbf{Oscillation amplitude-dependence of tip-surfaces forces.}
(\textbf{a}) A model tip-surface force is shown as a function of tip position
and velocity. The force is given by equation (\ref{eq:model-force}). The zoomed
inset shows the region close to the surface where the force is non-zero. The
black lines are tip orbits with different amplitudes. The force is only plotted
up to the maximum amplitude. (\textbf{b}-\textbf{e}) The corresponding
approach-retract force curves where the red curve is the force on approach and
the blue curve is the force during retract. The curves show a clear
amplitude-dependence of the dissipated energy. For small amplitudes, the force
curve is essentially given by the conservative part of the tip surface force.
For large amplitude, the dissipative interaction dominates and the approach and
retract curves change significantly.
\label{fig:force-disk}}
\end{figure}

The observation that the tip follows a simple orbit in the $z$-$\dot{z}$ phase
plane leads to a more compact description of the tip-surface force. Every point
in the phase plane is part of only one sinusoidal orbit with a specific
amplitude. Together with the assumption that the tip-surface force does not
depend on previous interaction cycles, every point in the phase plane can be
mapped to a unique history of the tip motion which starts at the maximum
distance from the sample surface during an oscillation cycle. Thus, the
dependence of the tip-surface force on the past tip trajectory can be
incorporated into the dependence on the tip position $z$ and velocity $\dot{z}$.
 Therefore, the interaction is completely described as a two dimensional
function in phase space. Such a function is depicted in Fig.
\ref{fig:force-disk}a. The plotted model force is a van der
Waals-Derjaguin-Muller-Toropov (DMT) force with an additional damping term that
depends exponentially on the tip position $z$. This force model is given by
\begin{equation}
F_{\mathrm{ts}}(z,\dot{z})=\begin{cases}
-\frac{HR}{6(a_{0}+h-z)^{2}}+\gamma_{0}\exp\left(\frac{z-h}{z_{\gamma}}\right)\dot{z} & z\leq h\\
-\frac{HR}{6a_{0}^{2}}+\frac{4}{3}E^{*}\sqrt{R}(z-h)^{3/2}+\gamma_{0}\exp\left(\frac{z-h}{z_{\gamma}}\right)\dot{z} & z>h
\end{cases}\label{eq:model-force}
\end{equation}
where $H$ is the Hamaker constant, $R$ is the tip radius, $a_{0}$ is the
inter-molecular distance, $\gamma_{0}$ is the damping constant, $z_{\gamma}$ is
the damping decay length and $E^{*}$ is the effective stiffness of the
tip-sample system. The assumed numerical values for these constants are shown in
table 1.

\begin{figure}[ht]
\includegraphics{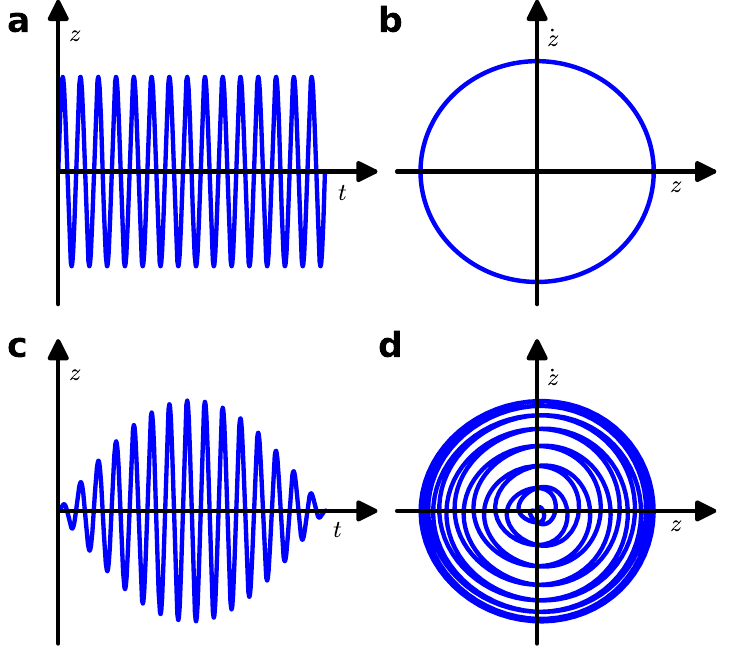}
\caption{\textbf{Tip motion for constant amplitude and modulated amplitude.}
(\textbf{a}) Tip motion with constant amplitude in the time domain. (\textbf{b})
The constant amplitude trajectory in  phase space  covers only one orbit and it
is not possible to reveal the complex dependence of the force on $z$ and
$\dot{z}$. (\textbf{c}) With modulated amplitude the tip performs a beat-like
motion in the time domain. The amplitude modulation is slow compared to the
frequency of the tip oscillation. At the lower turning points where the tip
interacts with the surface, the motion can be considered purely sinusoidal with
constant amplitude. (\textbf{d}) Over a full modulation period, a large area in
phase space is covered and the tip experiences the full $z$- and
$\dot{z}$-dependence of the tip-surface force.
\label{fig:tip-motion}}
\end{figure}

Tip motion with fixed amplitude is depicted in Fig. \ref{fig:tip-motion}a. The 
corresponding phase space trajectory in Fig. \ref{fig:tip-motion}b covers only
one simple orbit. Thus,  force measurement at only one amplitude does not reveal
the full nature of the tip-surface interaction as illustrated in Fig.
\ref{fig:force-disk}b-e where conventional force-distance curves for different
oscillation amplitudes are plotted. It is not possible to extract the
exponential dependence of the damping from only one measurement at fixed
amplitude. Several measurements with different amplitudes are required to cover
a larger region in the phase plane and gain a deeper insight into the nature of
the tip-surface force, especially regarding dissipative interactions which can
have very different physical origin.

\subsection{Force amplitude-dependence}
The tip-surface force could be considered as a time-dependent force acting on
the oscillator and the force could then be reconstructed from higher harmonics
of the tip motion\cite{Stark2002,Durig2000a}. However, equation
(\ref{eq:tip-motion}) implies that higher harmonics of the tip-motion are
insignificant. Indeed, their amplitudes are typically below the detection noise
floor and special force transducers\cite{Sahin2007}, high interaction
forces\cite{Stark2002} or highly damped environments\cite{Legleiter2006} are
required for their measurement. The high quality factor of the cantilevers
resonance amplifies only the Fourier components of the force near the resonance
frequency $\omega_{0}$, whereas higher harmonics are sharply attenuated. The
force Fourier component $\hat{F}_{\mathrm{ts}}(\omega_{0})$ can be determined
from knowledge of the calibrated cantilever linear response function
$\hat{\chi}$, the drive force $F_{\mathrm{drive}}$ and the equation of motion in
Fourier space
\begin{equation}
\hat{z}(\omega) = \hat{G}(\omega)\left( \hat{F}_{\mathrm{ts}}(\omega) + \hat{F}_{\mathrm{drive}}(\omega) \right) \label{eq:em-fourier}
\end{equation}
where $\hat{z}(\omega)$ is the Fourier transform of the tip motion
$z(t)$. With equation (\ref{eq:tip-motion}) the complex Fourier
component $\hat{F}_{\mathrm{ts}}(\omega_{0})$ can also be expressed
as two real-valued Fourier components that are in-phase and quadrature
to the motion
\begin{eqnarray}
F_{I} & = & \frac{1}{T}\int_{0}^{T}F_{\mathrm{ts}}\left(z(t),\dot{z}(t)\right)\left(\frac{z(t)-h}{A}\right)dt\label{eq:fi}\\
F_{Q} & = & \frac{1}{T}\int_{0}^{T}F_{\mathrm{ts}}\left(z(t),\dot{z}(t)\right)\left(\frac{\dot{z}(t)}{-\omega_{0}A}\right)dt\label{eq:fq}
\end{eqnarray}
where $T=2\pi/\omega_{0}$ is the oscillation period. $F_{I}$ and $F_{Q}$ can be
measured with high signal to noise ratio, providing a foundation for highly
accurate force reconstruction. $F_{I}$ and $F_{Q}$ are usually measured for
different probe heights $h$ above the
surface\cite{Durig2000,Sader2005,Holscher2006,Lee2006,Hu2008,Katan2009}.
However, their dependence on oscillation amplitude $A$ can also be used for
force reconstruction. Hence the name amplitude-dependence force spectroscopy
(ADFS).

The component $F_{I}$ is called the virial of the tip-motion and is only
affected by conservative tip-surface interactions\cite{Paulo2001}. Since the
conservative tip-surface force $F_{c}$ depends only on the tip position $z$, we
can rewrite equation (\ref{eq:fi}) as (see Methods) 
\begin{equation}
F_{I}(A)\approx-\frac{1}{2\pi A}\int_{0}^{A^{2}}\frac{F_{c}(-\sqrt{u})}{\sqrt{A^{2}-u}}du\label{eq:fi-rewritten}
\end{equation}
Equation (\ref{eq:fi-rewritten}) shows that $F_{I}(A)$ is the Abel integral
transform of the conservative part of the tip-surface
interaction\cite{Arfken2005}. The Abel transform has a unique inverse transform
with which we determine the conservative tip-surface force as
\begin{equation}
F_{c}(-z)=\frac{1}{z}\frac{d}{dz}\int_{0}^{z^{2}}\frac{\sqrt{\tilde{A}}F_{I}(\sqrt{\tilde{A}})}{\sqrt{z^{2}-\tilde{A}}}d\tilde{A}\label{eq:force-reconstruction}
\end{equation}
Thus, equation (\ref{eq:force-reconstruction}) enables force reconstruction at
fixed static tip-sample separation $h$, using only the oscillation amplitude
dependence of $F_{I}(A)$. 

For dissipative forces such a simple reconstruction is not possible since they
also depend on the tip velocity $\dot{z}$. However, the force component
$F_{Q}(A)$ can be interpreted as the amplitude-dependent energy loss per
oscillation cycle (see Methods),
\begin{equation}
E_{\mathrm{dis}}(A)=-2\pi A F_{Q}(A)\label{eq:energy-loss}
\end{equation}
The amplitude-dependence of the energy loss gives a signature of the dissipation
mechanism and equation (\ref{eq:energy-loss}) allows for the quantitative
exploration of the dissipative interaction at fixed probe height $h$, whereas
previously a slow change of $h$ was required\cite{Garcia2006}.

\subsection{Probing the tip-surface interaction}
To rapidly acquire the data for ADFS we use a multi-frequency lockin measurement
scheme\cite{Tholen2011} which combines the fast tip oscillation at the resonance
frequency $\omega_{0}$ of the cantilever with a slowly varying amplitude.
Different amplitude modulated drive schemes  have been considered
before\cite{Cuberes2001a,Jesse2007} but they allowed only for a qualitative
analysis of the tip-surface interaction. The beat-like tip motion resulting from
the drive signal is periodic with the beat frequency $\Delta\omega$ (Fig.
\ref{fig:tip-motion}c). The calibrated linear response function
$\hat{\chi}(\omega)$ of the cantilever is used to convert the measured tip
motion spectrum $\hat{z}(\omega)$ to the force signals $F_{I}$ and $F_{Q}$. Due
to the slow change of the oscillation amplitude, every oscillation cycle at the
fast frequency $\omega_{0}$ can be considered as an oscillation with fixed
amplitude. Over a full beat period a large region in phase space is covered
(Fig. 2d) and the complete dependence of $F_{I}(A)$ and $F_{Q}(A)$ on the
oscillation amplitude is revealed. The measurement time for complete $F_{I}(A)$
and $F_{Q}(A)$ is 2 ms in our experiments under ambient conditions. This high
acquisition speed makes it possible to combine high resolution surface imaging
with high accuracy force measurements in a single scan with normal scan rates.
To minimize feedback artifacts while scanning we evaluate $F_{I}(A)$ and
$F_{Q}(A)$ only while the oscillation amplitude increases.

Moreover, the high quality factor resonance guarantees a gradual change of the
oscillation amplitude, so that we can determine the tip-surface interaction
without deflection or amplitude jumps that are caused by instabilities which are
inherent to quasi-static and other dynamic force spectroscopy
methods\cite{Seo2008}.  Furthermore, we have not observed any  indication of
irregular motion or subharmonic generation when we  study the tip motion over
several beat periods while slowly approaching a  polystyrene surface.

\subsection{Numerical simulations}
We begin with numerical simulations of the equation of motion
(\ref{eq:equation-of-motion}) with the tip-surface force defined by equation
(\ref{eq:model-force}) and the numerical values for the force constants given in
table 1. We consider a standard cantilever in air with resonance frequency of
$f_{0}= \omega_{0} / (2\pi) =300\ \mathrm{kHz}$, quality factor of $Q=400$ and
stiffness of $k=40\ \mathrm{N/m}$. We apply a drive force with frequency $f_{0}$
and sinusoidally modulate the drive strength at a frequency of $\Delta f=500\
\mathrm{Hz}$. Without any tip-surface interaction, this drive force yields a
beat-like motion with a maximum amplitude of 25 nm. In the presence of the
tip-surface force, we let the system reach a steady state and then collect data
during two cycles of the amplitude modulation at a distance of 17 nm above the
surface.

\begin{figure}[ht]
\includegraphics{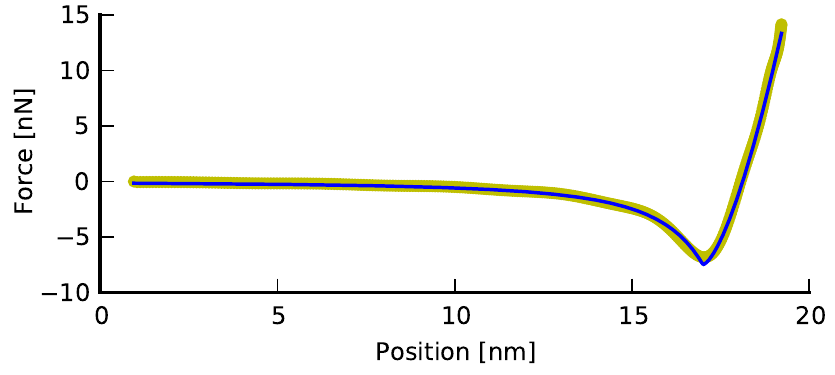}
\caption{\textbf{Conservative force reconstruction from simulated tip motion}.
The  reconstructed force curve (yellow) is excellent agreement with the actual
curve (blue) used in the simulation.
\label{fig:simulation}}
\end{figure}

From the simulated motion we extract $F_{I}(A)$ and reconstruct the conservative
tip-surface force $F_{c}$ using equation (\ref{eq:force-reconstruction}). The
reconstructed curve is shown in Fig. \ref{fig:simulation}, and it is in
excellent agreement with the actual force used in the simulations. At the sharp
transition point from the attractive to the repulsive regime small deviations
become visible in the order of 0.7 nN.

\subsection{Experimental results under ambient conditions} 
\begin{figure*}[ht]
\includegraphics{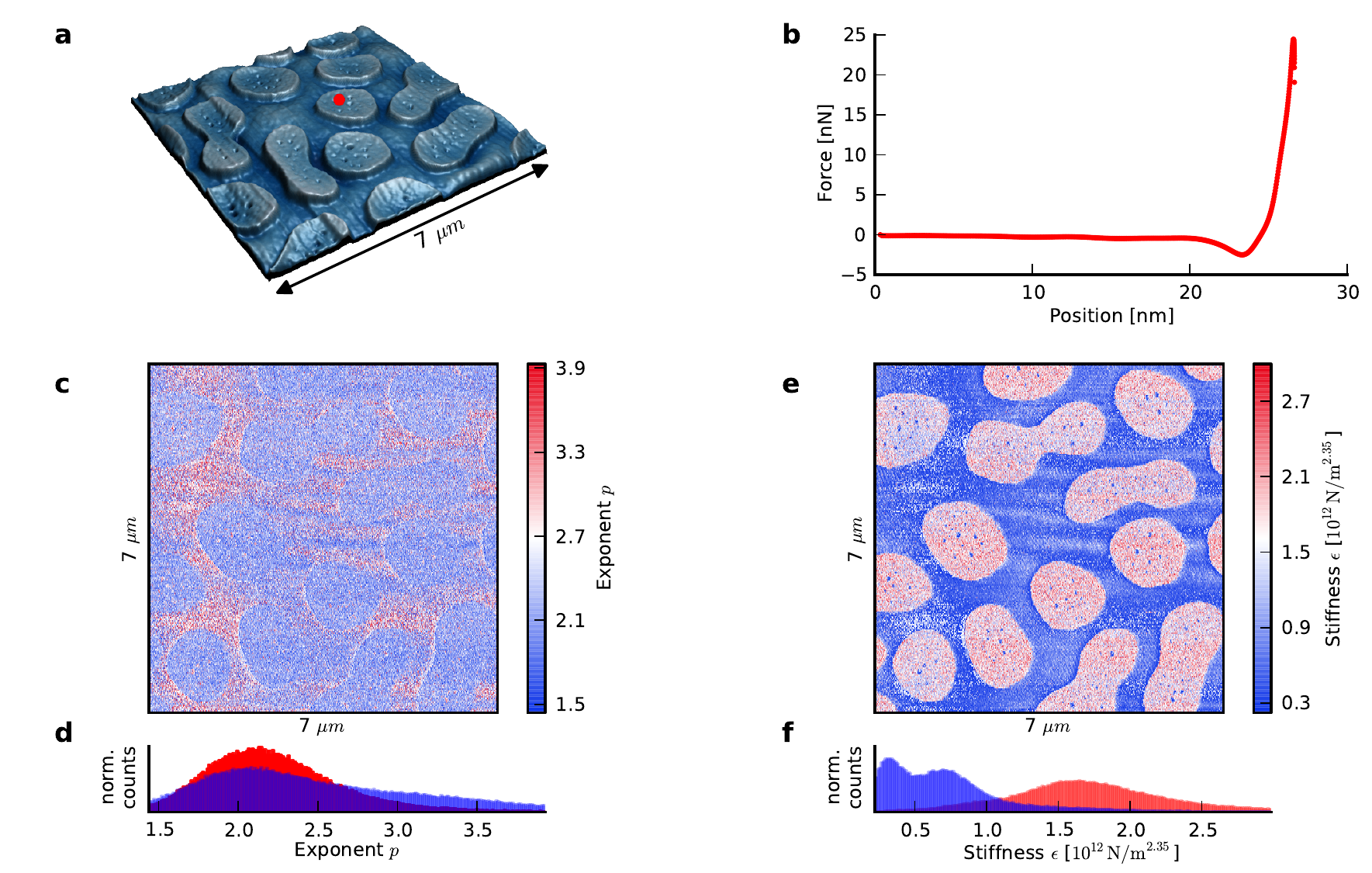}
\caption{\textbf{Conservative tip-surface force  measurements on a PS/PMMA
blend}. (\textbf{a}) The plateaus of the surface topography have a height of
circa 5 nm and are PMMA-rich, while the lower regions are PS-rich. (\textbf{b})
Reconstructed conservative force for the position marked with the red point in
(a). The force has an attractive minimum where the tip makes mechanical contact 
with the surface. Beyond this contact point the tip indents the surface and 
experiences a rapidly growing repulsive force. (\textbf{c}) Force reconstruction
in all points generates a map of the interaction exponent $p$ defined in equation
(\ref{eq:force-fit}). The map does not show a significant difference between the
PS and the PMMA domains. (\textbf{d}) Separate histograms of the interaction exponent
for the PS (blue) and PMMA (red) domains. (\textbf{e}) Map of the surface 
stiffness factor $epsilon$ defined in equation (\ref{eq:force-fit}). (\textbf{f}) 
Separate histograms of the stiffness factor for the PS (blue) and PMMA (red) 
domains. The PMMA-rich areas are a factor of three stiffer than the PS matrix.
\label{fig:conservative-force}}
\end{figure*}

In material science often the local mechanical properties of heterogeneous
polymer materials are of interest. One example of such a material is a blend of
polystyrene (PS) and poly(methyl methacrylate) (PMMA) with surface topography
shown Fig. \ref{fig:conservative-force}a. The different solubility of these
immiscible polymers causes the formation of domains with different height. The
higher structures in Fig. \ref{fig:conservative-force}a are PMMA-rich, while the
lower background matrix is PS-rich\cite{Walheim1997}. Scanning this sample we
collect data at 1024 points on each of 256 scan lines in 17 minutes. Without any
additional assumptions about the force or the motion, the conservative force
curve is determined using equation (\ref{eq:force-reconstruction}). A typical
conservative force curve at one image pixel is shown in Fig.
\ref{fig:conservative-force}b. The curve shows an attractive force close to the
surface with a sharp onset of repulsive force when the tip makes mechanical
contact with the surface. 

The ability to rapidly measure a conservative force curve with such high
accuracy at every image point enables a more detailed study of the local
nano-mechanical interaction. In the absence of adhesion the repulsive
conservative force is generally described in terms of a power law of the
form\cite{Sneddon1965}
\begin{equation}
F(z)=\epsilon(z-z_{\mathrm{min}})^{p}+F_{\mathrm{min}} \label{eq:force-fit}
\end{equation}
where $\epsilon$ is the local stiffness factor for a particular interaction
geometry and $z_{\mathrm{min}}$ is the position of the force minimum
$F_{\mathrm{min}}$. To minimize the adhesion during the mechanical contact
between the tip and the surface, we use a cantilever with small nominal radius
below 10 nm. A fit of the force model to the reconstructed force at every image
point reveals that the interaction exponent has a mean value of $2.35\ \pm\
0.3$, on both the PS- and PMMA-rich domains (see Fig. \ref{fig:conservative-force}c and d).
This observation suggests that the geometry of the contact is close to the ideal case
of a cone indenting a flat surface, where an exponent of $p=2.0$ is
expected\cite{Johnson1987}. We attribute this behavior to the actual conic shape
of the AFM tip. In contrast, traditional quantitative AFM methods assume the DMT
model geometry of a perfect sphere indenting a flat surface where an exponent of
$p=1.5$ is expected\cite{Derjaguin1975}. The measured interaction exponent
illustrates the difficulties of quantitative parameter extraction from AFM
measurements for which common approaches are based on various assumptions about 
surface adhesion, probe shape and interaction geometry. Furthermore, one should
note that the applicability of continuum mechanics models at the nanoscale is
still an open question\cite{Luan2005}.

Fixing the exponent $p$ for the whole surface, we generate a map of the
stiffness factor as shown in Fig. \ref{fig:conservative-force}e. Typically
dynamic AFM requires an order of magnitude difference in stiffness in order to
show image contrast. The high force accuracy of our measurement allows for the
resolution of a factor of three difference in stiffness as shown in Fig.
\ref{fig:conservative-force}f. The PMMA-rich domains are stiffer than the
PS-rich matrix, in agreement with previous measurements\cite{Sahin2008a} and in
agreement with the expected bulk modulus\cite{Brandrup2005}.

\begin{figure*}[ht]
\includegraphics{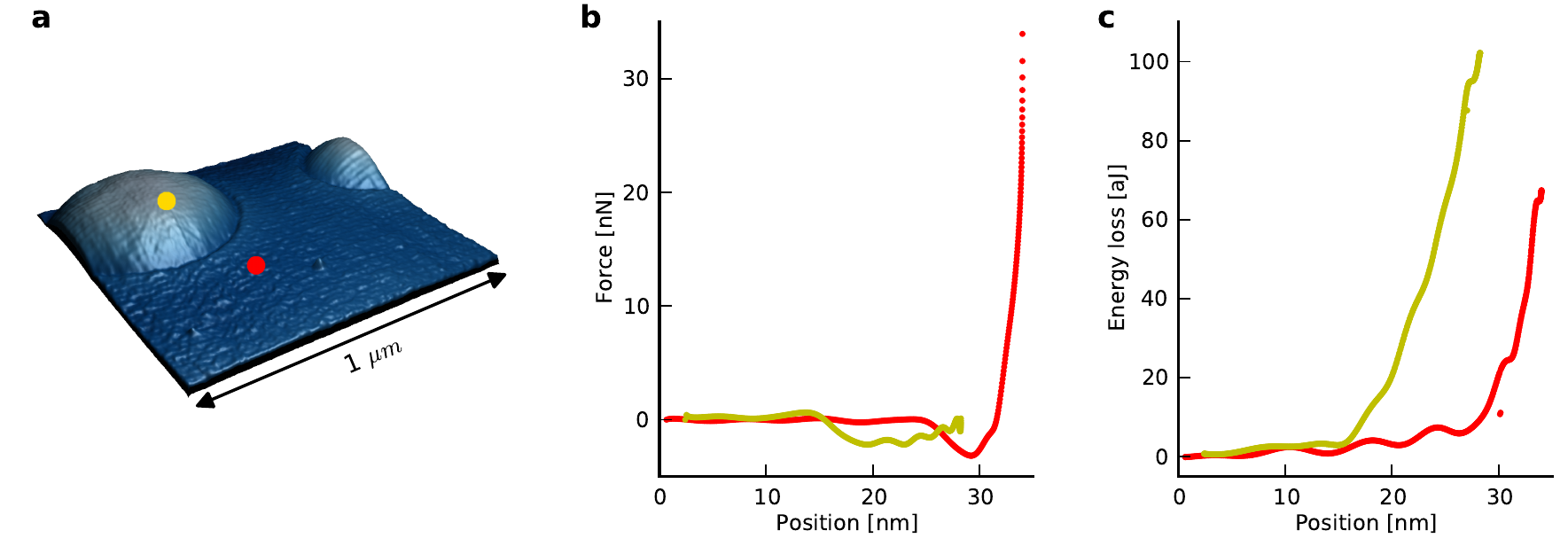}
\caption{\textbf{Conservative and dissipative tip-surface  interactions on a
PS/LDPE blend.}  (\textbf{a}) The droplet-like structures of the surface
topography are LDPE-rich.  (\textbf{b}) Reconstructed conservative tip-surface
force in the PS (red) and the LDPE (yellow)  domains. (\textbf{c}) Energy
dissipation curves in the PS (red) and the LDPE (yellow) domains.
\label{fig:dissipative-force}}
\end{figure*}

Local stiffness is a property typically extracted from the conservative part of
the surface force. However, Dissipative interactions are also present, which was
recognized early in the history of dynamic
AFM\cite{Cleveland1998,Paulo2001,Garcia2007}. These dissipative interactions are
far more complex as they depend not only on tip position $z$, but also on the
velocity and the history of tip motion. To get a signature of the dissipation
mechanism we use equation (\ref{eq:energy-loss}) to reconstruct the energy loss
per oscillation cycle as a function of the oscillation amplitude $A$. To
demonstrate this we scanned a heterogeneous blend of polystyrene and low density
polyethylene (LDPE). Figure \ref{fig:dissipative-force}a shows the surface
topography of the blend where LDPE aggregates to form droplet-like structures.
Figures \ref{fig:dissipative-force}b and \ref{fig:dissipative-force}c show
typical conservative force and energy loss curves reconstructed with ADFS on
each of the two domains. For both domains the dissipated energy in Fig.
\ref{fig:dissipative-force}c increases significantly in the repulsive region of
the conservative tip-surface force. For LDPE we see that both the dissipation
and the size of the interaction region are much larger than for PS. Moreover, on
LDPE the conservative force stays in the net attractive regime. We can interpret
these observations as resulting from greater positive and negative surface
deformation as the tip approaches and retracts above LPDE. This interpretation
is supported by the fact that the LDPE domains are not in the stiff glass phase
since the glass transition temperature of LDPE is below room
temperature\cite{Brandrup2005}.

\subsection{Liquid environments and high-frequency cantilevers}
The presented remarkably fast and accurate force measurements were performed
with standard cantilevers and under ambient conditions. However, one reason for
the wide-spread use of AFM is the ability to image  with a huge variety of
cantilevers in different environments such as liquids. We therefore consider the
applicability of ADFS in different operating regimes.

The ADFS data acquisition scheme is based on a slow modulation of the tip
oscillation amplitude which can be achieved by driving with two pure tones close
to the first flexural cantilever resonance. The spacing between the drive
frequencies determines the measurement bandwidth $\Delta f$ and in turn the
measurement time $T=1/\Delta f$. Higher measurement speed (scanning speed)
therefore requires larger spacing between the drive frequencies. The tip-surface
force generates new frequency components in the spectrum of the tip motion near
resonance, with spacing $\Delta f$\cite{Tholen2011}. These new components
contain the information for the ADFS reconstruction and usually the measurement
of approximately 20 frequency components is required for an accurate ADFS
reconstruction, as we have verified with simulations. The measurement therefore
requires a frequency band around resonance broad enough to collect these 20
frequency components with a signal-to-noise ratio (SNR) larger than 1. We call
this frequency band the resonant detection band.

The accuracy and speed of the ADFS technique improves with increasing width of
this resonant detection band. In this band the tip dynamics $z(t)$ are well
represented by the single harmonic oscillator which in Fourier space is
described by the linear response function $\hat{\chi}$ given by
\begin{equation}
\hat{\chi}(\omega)=\frac{1}{k}\frac{\omega_{0}^{2}}{\left(\omega_{0}^{2}-\omega^{2}\right)+i\frac{\omega_{0}\omega}{Q}}
\end{equation}
with the imaginary unit $i$. The linear response function is fully characterized
by  the parameters $k$, $\omega_0$, $Q$ which depend on the cantilever and the
environment. For a given force the response spectrum $\hat{z}$ is given by
equation (\ref{eq:em-fourier}) and the detector noise level determines the width
of the resonant detection band. 

\begin{figure*}[ht]
\includegraphics{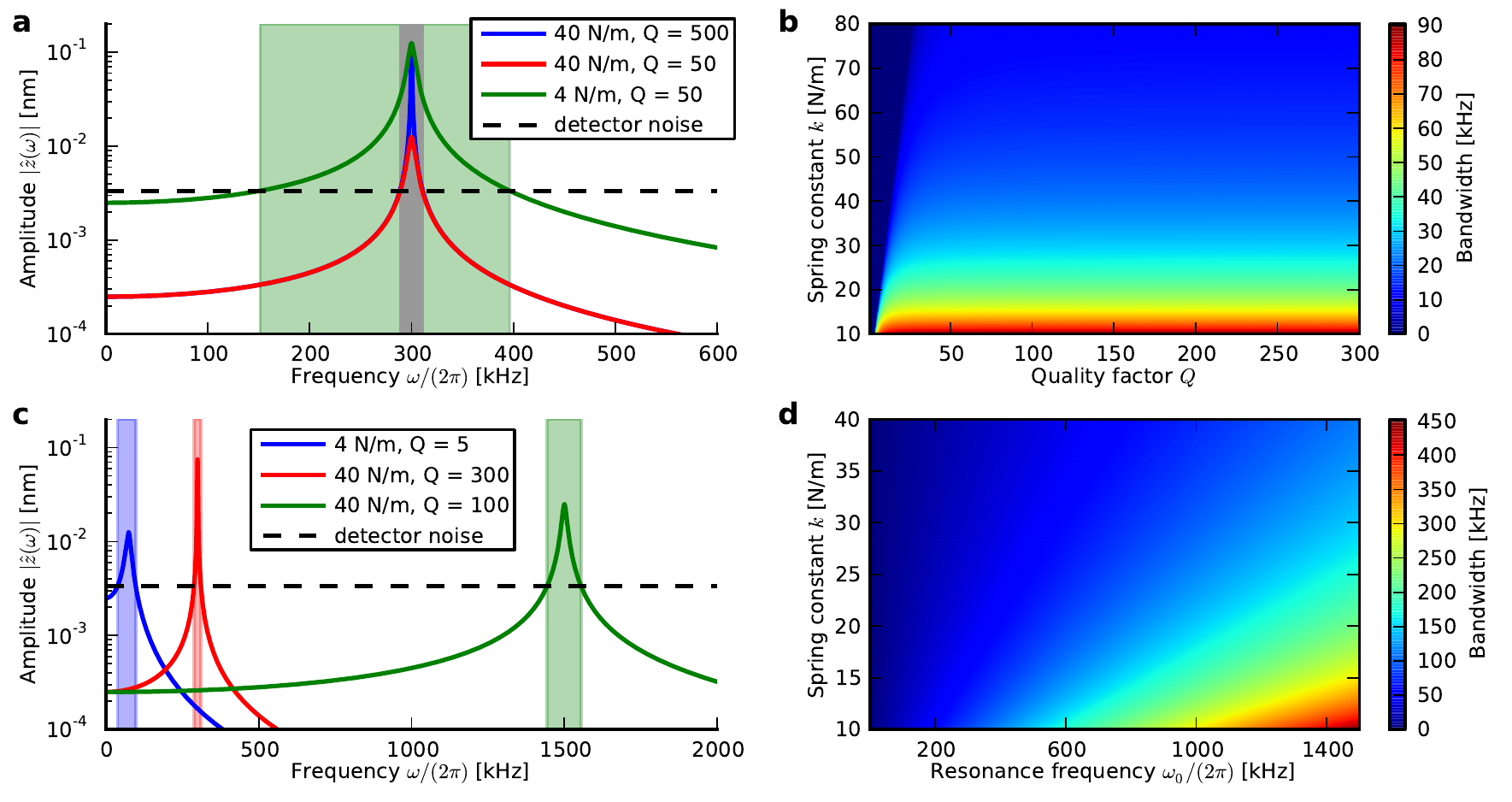}
\caption{\textbf{Dependence of the resonant detection bandwidth on  cantilever
parameters.} (\textbf{a}) Frequency-dependent cantilever response to a force of
0.01 nN. The noise level is determined for white detector noise of 150
$\mathrm{fm}/\sqrt{\mathrm{Hz}}$ and a measurement bandwidth of $\Delta f=500\
\mathrm{Hz}$. For an AFM cantilever with a resonance frequency of
$\omega_{0}=2\pi\cdot300\ \mathrm{kHz}$ the width of the resonant detection band
(shaded regions) is nearly independent of the quality factor but depends strongly on
the spring constant. (\textbf{b}) Continuous dependence of the resonant detection bandwidth
on the spring constant and quality factor for a  cantilever with a fixed
resonance frequency of $\omega_{0}=2\pi\cdot300\ \mathrm{kHz}$. Above a
threshold value of $Q$, the width of the resonant detection band is  
independent of the quality factor. (\textbf{c}) Frequency-dependent  cantilever
response to a force of 0.01 nN for typical cantilevers with resonance
frequencies of $\omega_{0}=2\pi\cdot75\ \mathrm{kHz}$ (blue), 300 kHz (red) and
1500 kHz (green). The width of the resonant detection band (shaded regions) increases for higher resonant
frequency and lower spring constant. (\textbf{d}) Continuous dependence of the
resonant detection bandwidth on the spring constant and resonant frequency for a
 cantilever with quality factor of $Q=300$.
\label{fig:bw}}
\end{figure*}

For the sake of discussion we take the detector noise level to be
frequency-independent  (white noise) with an amplitude of $150\
\mathrm{fm}/\sqrt{\mathrm{Hz}}$ and a measurement bandwidth of 500 Hz. In Fig.
\ref{fig:bw}a  we show the resonant detection band for a sinusoidal force of
strength 10 pN applied  to a cantilever having resonance frequency 300 kHz with
different spring constants  and quality factors. The red curve shows that a
significant reduction of the quality factor does not decrease the width of the
resonant detection band. On the other hand, reducing the cantilever spring
constant (for fixed $\omega_{0}$ and $Q$) does strongly increase the width of
the resonant detection band. In Fig. \ref{fig:bw}b the width of the resonant
detection band is plotted as a function of spring constant and quality factor
for fixed $\omega_{0}$. We note that for $Q$ larger than a low threshold value,
the width the resonant detection band is independent of $Q$ and increases
inversely with the spring constant $k$. Thus, decreasing $k$ appears
advantageous. However, the spring constant $k$ should not be too small since for
cantilevers with small spring constant the dependence of the motion on the
nonlinear tip-surface force in equation (\ref{eq:equation-of-motion}) becomes
stronger and the motion can become chaotic \cite{Hu2006,Jamitzky2006} . Smaller
$k$ also results in the generation of higher
harmonics\cite{Legleiter2006,Stark2010} and excitation of higher cantilever
eigenmodes\cite{Basak2007,Melcher2008a} which distort the motion so that it is
no longer nearly sinusoidal.

Another cantilever parameter that can be changed is the resonance frequency
$\omega_{0}$. Recently, low mass cantilevers with resonance frequency up to 2
MHz have become commercially available. As can be seen in Fig. \ref{fig:bw}c and
\ref{fig:bw}d the width of the resonant detection band increases linearly with
the resonance frequency (at fixed quality factor $Q$). One should note that
these low mass cantilevers are shorter than conventional cantilevers which
yields a bigger deflection angle for the same amplitude of tip motion, resulting
in greater responsivity of the optical lever system. This increased responsivity
lowers the detection noise floor, further increasing the resonant detection
bandwidth. Thus, the multifrequency data acquisition scheme used for ADFS
benefits strongly from this trend toward higher frequency cantilevers.

For operation in liquids often cantilevers with lower resonance frequencies and
lower spring constants are used. A small spring constant compensates for the
decrease of the resonant detection bandwidth due to the lower resonance
frequency. However, both spring constant and the resonance frequency should not
be too small in order to avoid distortion of the motion from higher
harmonics\cite{Legleiter2006,Stark2010} and higher
eigenmodes\cite{Basak2007,Melcher2008a}. Maximum oscillation amplitudes below 10
nm and high setpoint values also help to reduce these higher frequency
contributions, as has been demonstrated for torsional force
sensors\cite{Dong2009a}. Finally, we do not want the resonant detection band to
spread to too low frequencies where $1/f$-noise becomes dominant. To achieve
these goals, resonance frequencies of $\omega_{0}\approx2\pi\cdot75\
\mathrm{kHz}$ and spring constants of $k\approx4\ \mathrm{N/m}$ should be
sufficient. 

Our discussion thus far has focused on maximizing the resonant detection
bandwidth for a given detector noise. However, the SNR within this band will
have a contribution from the thermal noise force connected with the damping of the
cantilever motion in the surrounding medium. The thermal noise force is
independent of frequency and its magnitude is given by the fluctuation
dissipation theorem as 
\begin{equation}
F_{\mathrm{th}}=\sqrt{2k_{B}T\left(\frac{k}{\omega_{0}Q}\right)\Delta f}.
\end{equation}
When going from air to liquid for fixed spring constant $k$, the thermal noise
force will significantly increase due to the reduction in $\omega_{0}$ and $Q$.
If the thermal noise dominates over the detector noise, it can be shown that the
SNR is independent of the spring constant $k$\cite{Platz2012a}.

\section{Discussion}
We demonstrated accurate and high resolution measurements of the conservative
force and the dissipated energy while scanning. A direct analysis of the
measured tip motion allows us to gain insight detailed information about the
nature of the tip-surface interaction and dissipation mechanisms without
extensive modeling based on idealized assumptions about the geometry of the tip
and the surface. Such insight is often necessary to understand the contrast
mechanism in AFM images.

ADFS requires only a narrow detection band whose width is basically independent
of the cantilever quality factor. Therefore,  ADFS can also be used for
investigating biological samples in highly damped liquid environments with
present-day noise detection limits. However, the tip motion has to be
approximately sinusoidal on the level of single oscillation cycles. 

Furthermore, ADFS is compatible with the current trend in AFM toward smaller,
higher frequency probes\cite{Li2007,Arlett2011,Mininni2011} since the width of
the detection band increases linearly with the cantilever resonance  frequency.
With future generations of  high frequency force sensors we envision that ADFS
will enable high-resolution force-volume measurement at video rate. 

The general idea of exploring amplitude-dependence can also be extended to other
modes of AFM like frequency-modulated AFM, torsional shear force microscopy or 
frictional force microscopy. Beyond AFM, the ADFS concept may inspire the
development of new sensing schemes, as the basics concepts presented here are
applicable to many types of measurements which exploit the enhanced sensitivity
of a high-quality-factor resonator.

\section{Methods}

\subsection{Force reconstruction}

For a single drive frequency the tip motion is approximately sinusoidal as given
by equation (\ref{eq:tip-motion}) and the  corresponding tip velocity becomes
\begin{equation}
\dot{z}(t) \approx -\omega_{0}A\sin(\omega_{0}t+\phi)\label{eq:tip-velocity}
\end{equation}
As the tip oscillates it experiences the surface force which is described as a
function of tip position $z$ and tip velocity $\dot{z}$. The force is usually
decomposed into a conservative part and into a dissipative or non-conservative
part as in equation (\ref{eq:force-decomposition}). The conservative force is
considered to be a function of only the tip position $z$ while the dissipative
part depends also on the tip  velocity $\dot{z}$, such that the dissipative
force is anti-symmetric with respect to the velocity
\begin{equation}
F_{\mathrm{nc}}(z,-\dot{z})=-F_{\mathrm{nc}}(z, \dot{z})\label{eq:f-nc-antisymmertry}
\end{equation}

If the tip-motion contains only one frequency, only two real-valued  Fourier
components of the time-dependent force between tip and surface are measurable.
With an appropriate shift of coordinates we set $h=0$ and with equations (\ref{eq:force-decomposition}), (\ref{eq:tip-motion}),  
(\ref{eq:tip-velocity}) and (\ref{eq:f-nc-antisymmertry}) these components given
in equation (\ref{eq:fi}) and (\ref{eq:fq}) can be written as
\begin{eqnarray}
F_{I}(A) & = & \frac{1}{T}\int_{0}^{T}F_{\mathrm{c}}\left(A\cos(\omega_{0}t)\right)\cos(\omega_{0}t)dt\label{eq:fi-int-2}\\
F_{Q}(A) & = & \frac{1}{T}\int_{0}^{T}F_{\mathrm{nc}}\left(A\cos(\omega_{0}t),-\omega_{0}A\sin(\omega_{0}t)\right)\sin(\omega_{0}t)dt\label{eq:fq-int-2}
\end{eqnarray}
We note that $F_{I}$ and $F_{Q}$ are functions of the oscillation amplitude and
that $F_{I}$ depends only on the conservative part of the tip-sample interaction
whereas $F_{Q}$ is only affected by the dissipative interaction.

With the substitution $z'=A\cos(\omega_{0}t)$ we rewrite equation
(\ref{eq:fi-int-2}) as
\begin{equation}
F_{I}(A)=\frac{1}{\pi}\int_{-A}^{A}F_{\mathrm{c}}(z')\frac{z'/A}{\sqrt{A^{2}-{z'}^{2}}}dz'. \label{eq:fi-int-z}
\end{equation}
The tip-surface interaction is localized to a small region close to the surface
which is small compared to the oscillation amplitude. With this assumption and
the substitution $u={z'}^{2}$ one obtains equation (\ref{eq:fi-rewritten}) Now,
we define $\tilde{A}=A^{2}$, 
$\tilde{F}_{\mathrm{c}}(u)=F_{\mathrm{c}}\left(-\sqrt{u}\right)$ and
$\tilde{F}_{I}(\tilde{A})=-2\pi F_{I}\left(\sqrt{\tilde{A}}\right)$ and rewrite
equation (\ref{eq:fi-rewritten}) as
\begin{equation}
\tilde{F}_{I}(\tilde{A})=\int_{0}^{\tilde{A}}\frac{\tilde{F}_{\mathrm{c}}(u)}{\sqrt{\tilde{A}-u}}du\label{eq:fi-int-abel}
\end{equation}
Equation (\ref{eq:fi-int-abel}) reveals that $\tilde{F}_{I}(\tilde{A})$ is the
Abel transform of $\tilde{F}_{\mathrm{c}}(u)$. Similar integrals have been
studied before\cite{Durig2000,Sader2005,Holscher2006} but they were never
considered as a function of amplitude. The Abel transform has a unique
inverse\cite{Arfken2005} which enables the solution of equation
(\ref{eq:fi-int-abel}) for the force $\tilde{F}_{\mathrm{c}}(u)$
\begin{equation}
\tilde{F}_{\mathrm{c}}(u) = \frac{1}{\pi}\frac{d}{du}\int_{0}^{u}\frac{\tilde{F}_{I}(\tilde{A})}{\sqrt{u-\tilde{A}}}d\tilde{A}\label{eq:fc-u}
\end{equation}
which can readily be rewritten as equation (\ref{eq:force-reconstruction}). For
experimental data equation (\ref{eq:force-reconstruction}) has to be evaluated
numerically. To improve the numerical stability we perform the substitution
$y^{2}=z^{2}-\tilde{A}$ which removes the square root singularity,
\begin{equation}
F_{\mathrm{c}}(z)=-\frac{2}{z}\frac{d}{dz}\int_{0}^{z}\sqrt{z^{2}-y^{2}} F_{I}\left(\sqrt{z^{2}-y^{2}}\right)dy\label{eq:fc-y-subst}
\end{equation}
With equation (\ref{eq:fc-y-subst}) we can reconstruct the conservative
tip-surface interaction without any assumption regarding its functional
representation.

The non-conservative interaction can be characterized by the energy
dissipated per oscillation cycle,
\begin{equation}
E_{\mathrm{dis}}=\oint F_{\mathrm{nc}}(z,\dot{z})dz=\int_{0}^{T}F_{\mathrm{nc}}\left(z(t),\dot{z}(t)\right)\dot{z}(t)dt\label{eq:E-dis-int}
\end{equation}
With the equations (\ref{eq:tip-velocity}) and (\ref{eq:E-dis-int}) 
we can rewrite the integral equation (\ref{eq:fi-int-2}) in terms
of the dissipated energy $E_{\mathrm{dis}}$,
\begin{equation}
F_{Q}(A)=-\frac{1}{2\pi A}\int_{0}^{T}F_{\mathrm{nc}}\left(z(t),\dot{z}(t)\right)\dot{z}(t)dt=-\frac{E_{\mathrm{dis}}}{2\pi A}
\end{equation}
which allows for the study of the energy dissipation from the amplitude-dependence
of $F_{Q}(A)$ as in equation (\ref{eq:energy-loss}).

\subsection{Numerical methods}
The numerical simulations have been performed using the CVODE adaptive step-size
integrator\cite{Hindmarsh2005}. Special care was taken to treat the non-smooth
transition between attractive and repulsive forces at $z=0$ in the force defined
by equation (\ref{eq:model-force}).

\subsection{Experimental methods}
Polystyrene ($M_w = 280\ \mathrm{kDa}$), poly(methyl methacrylate) ($M_w= 120\
\mathrm{kDa}$) and toluene were obtained from Sigma-Aldrich and used as
purchased. The polymers were dissolved in toluene at a concentration of 0.52
wt\% and spin-cast on a silicon substrate. The sample was scanned with a
Multimode AFM and a Nanoscope 4 controller (Bruker). The PS/LDPE sample (Bruker)
was scanned with a Dimension 3100 AFM and a Nanoscope 3a controller (Bruker).
Both AFM systems were used together with a signal access module and a
multi-frequency lockin analyzer (Intermodulation Products AB, IMP 2-32) for
drive synthesis and signal analysis. All scans have been performed with
Tap300Al-G probes (Budget Sensors) which were calibrated with a non-invasive
thermal method\cite{Higgins2006}.

\section{Acknowledgments}
The authors acknowledge financial support from the Swedish Research Council (VR)
and the Swedish Government Agency for Innovation Systems (VINNOVA).

\section{Author Contributions}
D.P. developed the measurement concept and the theory. D.P. fabricated the 
PS/PMMA sample. All authors contributed to the experimental setup. D.P. and D.F.
performed the measurements. D.P. analyzed the data. D.P. and D.B.H wrote the
manuscript. All authors discussed and contributed to the manuscript.

\section{Competing Financial Interests}
Two patent applications ("Intermodulation Scanning Force Spectroscopy", 
PCT/EP2008/066247 and "Intermodulation Lock-in", USPTO 13098597) on the 
multifrequency measurement techniques have been filed by the authors.

\section{Figure Legends}

\clearpage

\section*{Tables}
\begin{table}[ht]
\caption{\textbf{Force parameters.} Values of the parameters used in the numerical simulations and for the force plotted in Fig. 1}
\begin{tabular*}{\hsize}{@{\extracolsep{\fill}}ll}
Force parameter                   &    Value  \cr
\hline
Hamaker constant $H$              &    $3.28 \times 10^{-17}\ \mathrm{J}$\cr
Tip radius $R$                    &    $10\ \mathrm{nm}$\cr
Inter-molecular distance $a_{0}$  &    $2.7\ \mathrm{nm}$\cr
Effective stiffness $E^{*}$       &    $1.50\ \mathrm{GPa}$\cr
Damping factor $\gamma$           &    $2.2 \times 10^{-7}\ \mathrm{\frac{kg}{s}}$\cr
Damping decay length $z_{\gamma}$ &    $1.5\ \mathrm{nm}$\cr
\end{tabular*}
\end{table}

\end{document}